\documentclass[pdflatex,sn-mathphys]{sn-jnl}

\jyear{2021}%

\theoremstyle{thmstyleone}%
\theoremstyle{thmstyletwo}%

\theoremstyle{thmstylethree}%

\usepackage{braket}
\usepackage{siunitx}
\usepackage{nicefrac}
\usepackage{graphicx}
\usepackage{dcolumn}
\usepackage{bm}
\usepackage{xcolor}
\usepackage[makeroom]{cancel}
\usepackage{mathtools}
\usepackage{comment}

\usepackage{enumitem}
\usepackage{interval}
\usepackage{amsmath,amssymb}
\usepackage{interval}
\usepackage[flushleft]{threeparttable}
\usepackage{booktabs}
\usepackage{siunitx}
\usepackage{textgreek}
\usepackage{datetime}
\usepackage{lineno}

\newdateformat{monthyeardate}{%
\monthname[\THEMONTH] \THEYEAR}

\begin{document}

\title[Nuclear Excitation by Muon Capture]{Nuclear Excitation by Muon Capture}

\author*[1]{Simone Gargiulo}\email{simone.gargiulo@epfl.ch}%
\author[2]{Ming Feng Gu}
\author[1]{Fabrizio Carbone}
\author*[1]{Ivan Madan}\email{ivanmadan@gmail.com}%

\affil[1]{%
 Institute of Physics (IPhys), Laboratory for Ultrafast Microscopy and Electron Scattering (LUMES), École Polytechnique Fédérale de Lausanne (EPFL), Lausanne 1015 CH, Switzerland}%
\affil[2]{%
Space Science Laboratory, University of California, Berkeley, CA 94720, USA}%

\abstract{ 
Efficient excitation of nuclei via exchange of a real or virtual photon has a fundamental importance for nuclear science and technology development. Here, we present a new mechanism of nuclear excitation based on the capture of a free muon into the atomic orbits (NEμC). The cross section of such a new process is evaluated using the Feshbach projection operator formalism and compared to other known excitation phenomena, i.e. photo-excitation and nuclear excitation by electron capture (NEEC), showing up to ten orders of magnitude increase in cross section. NEμC is particularly interesting for MeV excitations that become accessible thanks to the stronger binding of muons to the nucleus. The binding energies of muonic atoms have been calculated introducing a state of the art modification to the Flexible Atomic Code (FAC). An analysis of an experimental scenarios in the context of modern muon production facilities shows that the effect can be detectable for selected isotopes. The total probability of NEμC is predicted to be $ P \approx \SI{1e-6}{}$ per incident muon in a beam-based scenario. Given the high transition energy provided by muons, NEμC can have important consequences for isomer feeding and particle-induced fission.
}


\maketitle
Manipulating nuclear transitions is a highly desirable goal due to its implications in the energy sector \cite{aprahamian2005long, Walker1999a,collins1999accelerated,litz2004controlled,collins2004accelerated,carroll2013nuclear,vanacore2018attosecond,Madan2020}. 
Long-lived nuclear excitations, formally isomers, have lifetimes that are sometimes comparable to the age of the universe and have a potential to release hundreds of megajoules of energy stored in few cubic centimeter. The former aspect is crucial in designing new energy storage solutions: long-duration has been suggested to be the key driver towards decarbonized future \cite{sepulveda2021design}. 
Unfortunately, an efficient process to excite and control the lifetime of isomers is currently lacking. 

\begin{figure}[h]
\centering
\includegraphics[width=1\linewidth]{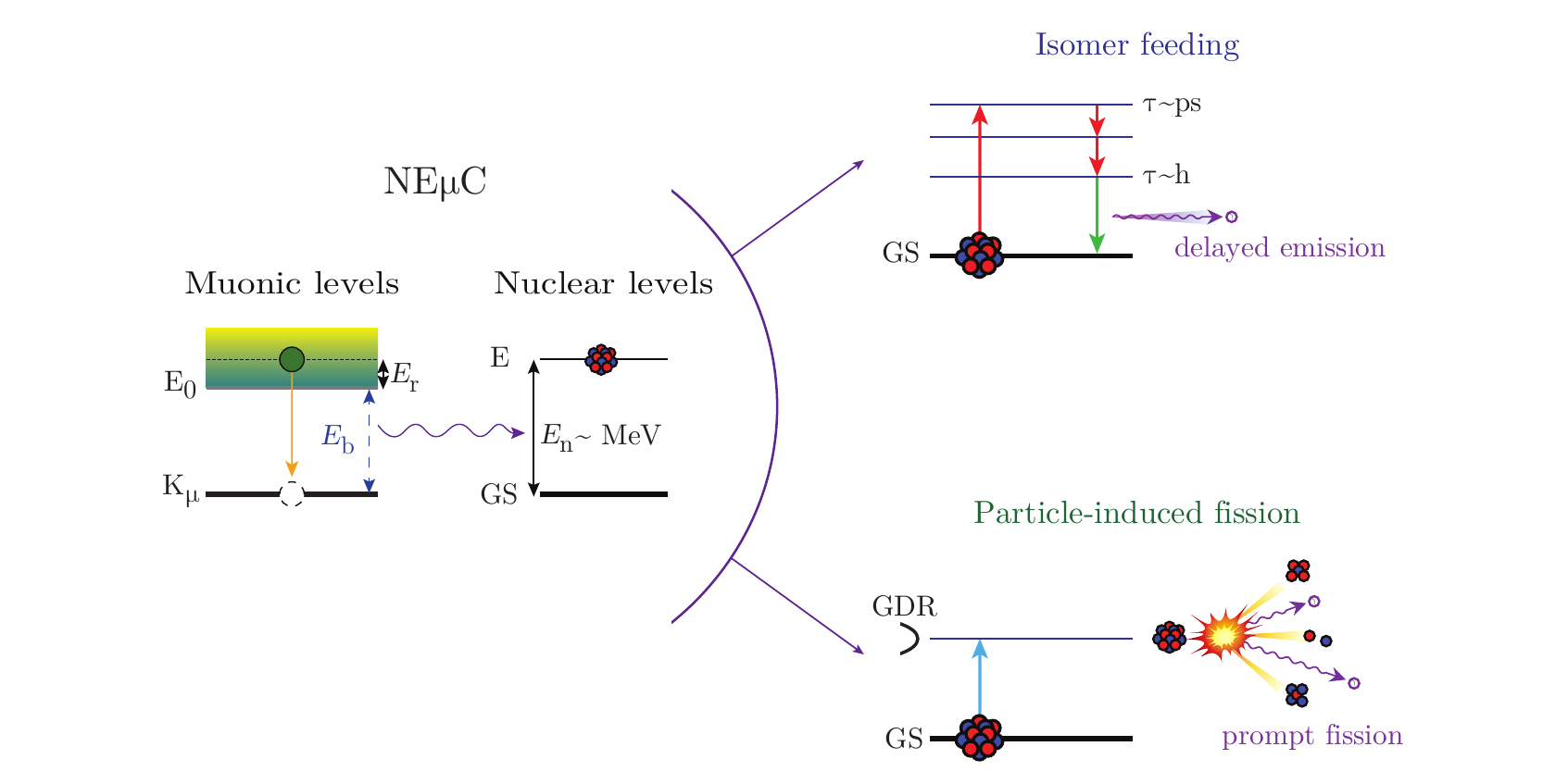}
\caption{Nuclear excitation by muon capture (NE\textmu C): the capture of a free muon leads to a resonant excitation of the nucleus. The excited nucleus can subsequently decay towards lower levels reaching a long-lived state, i.e. isomer feeding. Another possibility is present if the excitation of the nucleus is in resonance with the giant dipole (GDR) or quadrupole resonances (GQR), and the latter is above the fission barrier: the nucleus can undergo a prompt fission induced by NE\textmu C.}
\label{fig:Perspective}
\end{figure}

Nuclear levels, in general, are not easily accessible: they  often have high spin with respect to the ground state, the resonances are very narrow and predominantly in the \SI{}{\mega\electronvolt} range, in which no high-intensity monochromatic light sources exist yet. Alternatively to direct excitation via photon absorption, few other secondary electro-magnetic processes exist - such as Coulomb excitation \cite{pieper1957coulomb}, nuclear excitation upon electron capture (NEEC) \cite{Goldanskii1976,Chiara2018}  or transition (NEET) \cite{morel2004nuclear,Morel2010} and excitation upon muon cascade \cite{osti_4047420,Hoehn1979}. Here we discuss a new electro-nucleus excitation mechanism that presents one of the highest excitation cross sections: the nuclear excitation by muon capture (NE\textmu C).
The high energy transferred to the nucleus expands the range of isotopes suitable for the process and makes NE\textmu C a relevant process for muon-induced fission \cite{wheeler1949some,zaretski1961theory,oberacker1993muon} and for feeding of long-lived isomers \cite{aprahamian2005long}, as shown in Fig. \ref{fig:Perspective}.

\noindent
It is instructive to compare electronic and muonic electro-nucleus processes. Excitation upon muon cascade \cite{jacobsohn1954nuclear,wiletsexcitation} and NEET occur as a result of transition between bound muonic and electronic orbitals respectively. The two processes have been experimentally observed \cite{osti_4047420,Hoehn1979} and are considered to be well established. The NEEC process is believed to be observed in a single experiment \cite{Chiara2018,guo2021possible,chiara2021reply} and its cross section differs from a theoretical estimate by nine orders of magnitude \cite{Wu2019}. To our knowledge, a muonic analogue of NEEC has never been proposed and in this article we investigate this possibility theoretically, underlining key differences with the known processes.

Similarly to NEEC and contrary to NEET and muon cascade excitation, NE\textmu C describes capture of a \textit{free} lepton in a corresponding atomic orbital, and is thus not constrained by the restriction of matching the transition energies between \textit{bound} atomic and nuclear levels. Since both NEEC and NE\textmu C depend on the interaction between nuclear and atomic environment, tight muon orbits are expected to provide higher nuclear excitation cross section than their electronic counter parts.
In particular, here we report findings of NE\textmu C integrated cross section up to \SI{1.82e5}{\barn \electronvolt}, that is five orders of magnitude higher than any corresponding NEEC cross section reported so far \cite{Palffy2006,palffy2007quantum,Rzadkiewicz2019,wu2019x,gargiulo2021nuclear}.
To evaluate the NE\textmu C cross section we used the advanced theory based on the Feshbach projection operator formalism developed by A. Pálffy for the NEEC process and presented in Refs. \cite{Gagyi-Palffy06,Palffy2006}.
In this context, the NE\textmu C rate for an electric transition can be written in muonic atomic units as: 

\begin{eqnarray}
 Y_\mathrm{n}^{(e)L} &=& \dfrac{1}{4\pi\alpha} \dfrac{4\pi^2 \rho_i}{(2L+1)^2 } {B\! \uparrow} (EL) (2 j_b +1)\nonumber \\
  & &\times \sum_k \lvert\widetilde{R}_{L,k_b,k}\rvert^2 C(j_b\ L\ j; 1/2\ 0\ 1/2)^2\ ,
\end{eqnarray}
\noindent
where \textit{B}$\uparrow$(EL) is the reduced transition probability of the L\textsuperscript{th} multipolar transition, $C (j_1\ j_2\ j;\ m_1\ m_2\ m)$ is the Clebsch-Gordan coefficient, $j_\mathrm{b}$ and $j$, are  the total angular momentum, while $k_\mathrm{b}$ and $k$ are the Dirac angular momentum of the bound and free muon respectively. $\widetilde{R}_{L,k_b,k}$ is the radial integral that depends on the muon bound and free wavefunctions, which are obtained as solutions of the Dirac equations for a specific atomic configuration using the modified version of FAC \cite{gu2008flexible}. 

The NE\textmu C cross section can be expressed as: 

\begin{equation}
\sigma_\mathrm{NE\text{\textmu} C} = 2\pi^2 \lambda_\mathrm{\text{\textmu}}^2\ Y_\mathrm{n}\ \dfrac{\nicefrac{\Gamma_\mathrm{r}}{2\pi}}{(E-E_\mathrm{r})^2+\nicefrac{\Gamma_\mathrm{r}^2}{4}}\ ,
\label{eq:NEEC_CS}
\end{equation}
\noindent
where $\lambda_\text{\textmu}$ is the free muon wavelength and $E_\mathrm{r}$ the resonance energy. The integration of the cross sectioon over the continuum energies, considering that $\Gamma_\mathrm{r}<<E_\mathrm{r}$, gives the so-called resonance strength:
\begin{equation}
S_\mathrm{NE\text{\textmu} C} = \int \sigma_\mathrm{NE\text{\textmu} C}(E)\ \mathrm{d}E =  2\pi^2 \lambda_\mathrm{\text{\textmu}}^2\ Y_\mathrm{n}\ .
\label{eq:NEEC_RS}
\end{equation}
\noindent

\begin{figure}[h]
\centering
\includegraphics[width=\linewidth]{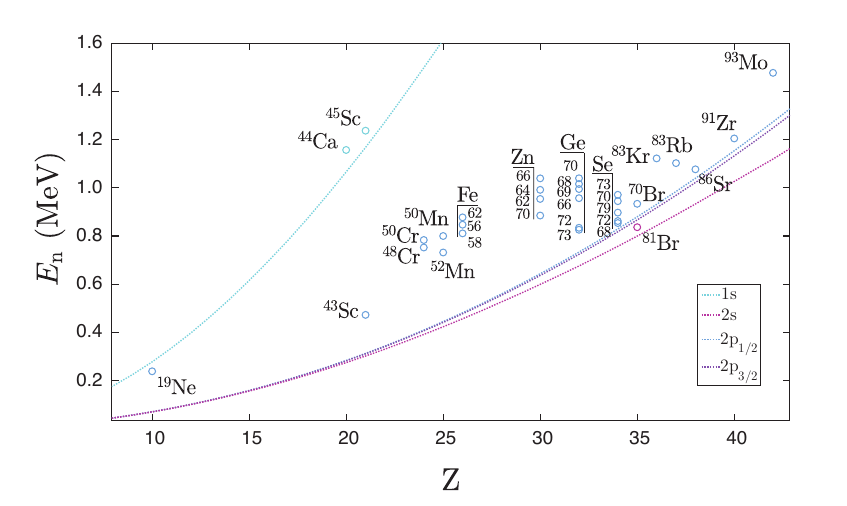}
\caption{Isotopes matching the search criteria in case of E2 transition. Only the nuclear transitions with a $B\! \downarrow\!(E2)>10 $ W.u. have been included in the plot. The color of the markers indicate the closest shell allowing NE\textmu C. Vertical black lines group several isotopes of the same element.}
\label{fig:Isotopes}
\end{figure}

For NE\textmu C to be possible the nuclear transition energy ($E_\mathrm{n}$) has to be larger than the muon binding energy ($ E_\mathrm{b}$), and the free muon energy ($E_\mathrm{r}$) has to match their difference (i.e., $E_\mathrm{r}=E_\mathrm{n} - E_\mathrm{b}$). This condition defines the search for the nuclear transitions that can be excited by the NE\textmu C mechanism.
In Fig. \ref{fig:Isotopes}, we plot the muonic binding energies for K and L shells, calculated with the Flexible Atomic Code (FAC) \cite{gu2008flexible} modified for muonic atoms (see Methods), and nuclear excited levels that satisfy the above criteria for $E_\mathrm{r}$  up to \SI{0.4}{\mega\electronvolt} above the corresponding muonic levels. For the sake of the presentation we only show the E2 transition, which are generally the strongest. The table \ref{tab:Calculations} reports the NE\textmu C rates and resonance strengths for selected isotopes together with nuclear transition energy and required energy of the free muon, including several E1 transitions. 

\noindent

\begin{table*}[!hbtp]
\centering
\caption{\label{tab:Calculations} Resonance strengths for the isotopes highlighted by the searching criteria for E1 and E2 transitions and capture in the K and L shells. Isotopes are ordered with respect to the mass number.} 
\begin{threeparttable}
\begin{tabular}{S|S|S|S|S|S|S} \toprule
{Isotope}&{L$^\mathrm{th}$}&{$E_{\mathrm{n}}$ (keV)}&{$nl_j$}&{$E_{\mathrm{r}}$ (keV)}&{$Y_{\mathrm{NE\text{\textmu} C}}$ (1/s)} &{$S_{\mathrm{NE\text{\textmu} C}}$ (b eV)}\\ \midrule 
{$^{11}$Be}&{E1}&{320.04}&{$\rm 1s_{1/2}$}&{275.54}&{\SI{1.39e+14}{}}&{\SI{12.09}{}}\\
{$^{19}$Ne}&{E2}&{238.27}&{$\rm 2p_{3/2}$}&{168.21}&{\SI{1.16e+14}{}}&{\SI{16.58}{}}\\
{$^{43}$Ca}&{E1}&{1394.47}&{$\rm 1s_{1/2}$}&{329.02}&{\SI{4.95e+10}{}}&{\SI{3.62e-3}{}}\\
{$^{44}$Ca}&{E2}&{1157.02}&{$\rm 1s_{1/2}$}&{92.20}&{\SI{4.14e+12}{}}&{\SI{1.08}{}}\\
{$^{45}$Sc}&{E2}&{1236.70}&{$\rm 1s_{1/2}$}&{69.08}&{\SI{2.84e+12}{}}&{\SI{0.99}{}}\\
{$^{48}$Cr}&{E2}&{752.19}&{$\rm 2s_{1/2}$}&{360.43}&{\SI{4.69e+13}{}}&{\SI{3.13}{}}\\
{$^{48}$Cr}&{E2}&{752.19}&{$\rm 2p_{1/2}$}&{342.84}&{\SI{2.63e+16}{}}&{\SI{1.84e3}{}}\\
{$^{48}$Cr}&{E2}&{752.19}&{$\rm 2p_{3/2}$}&{345.88}&{\SI{5.15e+16}{}}&{\SI{3.58e3}{}}\\
{$^{52}$Mn}&{E2}&{731.66}&{$\rm 2p_{1/2}$}&{287.05}&{\SI{9.28e+15}{}}&{\SI{771.14}{}}\\
{$^{68}$Se}&{E2}&{853.75}&{$\rm 2s_{1/2}$}&{92.94}&{\SI{3.52e14}{}}&{\SI{90.87}{}}\\
{$^{68}$Se}&{E2}&{853.75}&{$\rm 2p_{1/2}$}&{24.54}&{\SI{1.42e+17}{}}&{\SI{1.38e5}{}}\\
{$^{68}$Se}&{E2}&{853.75}&{$\rm 2p_{3/2}$}&{36.18}&{\SI{2.75e+17}{}}&{\SI{1.82e5}{}}\\
{$^{73}$Ge}&{E2}&{825.8}&{$\rm 2p_{1/2}$}&{92.66}&{\SI{3.81e+16}{}}&{\SI{9.85e3}{}}\\
{$^{73}$Ge}&{E2}&{825.8}&{$\rm 2p_{3/2}$}&{101.84}&{\SI{7.41e+16}{}}&{\SI{1.75e4}{}}\\
{$^{81}$Br}&{E2}&{836.8}&{$\rm 2s_{1/2}$}&{37.22}&{\SI{1.43e+14}{}}&{\SI{91.82}{}}\\
{$^{86}$Sr}&{E2}&{1076.7}&{$\rm 2p_{3/2}$}&{54.77}&{\SI{9.52e+15}{}}&{\SI{4.17e3}{}}\\
{$^{91}$Zr}&{E2}&{1204.8}&{$\rm 2p_{1/2}$}&{51.18}&{\SI{1.32e+16}{}}&{\SI{6.19e3}{}}\\
{$^{91}$Zr}&{E2}&{1204.8}&{$\rm 2p_{3/2}$}&{72.26}&{\SI{2.56e+16}{}}&{\SI{8.49e3}{}}\\
{$^{93}$Mo}&{E2}&{1477.2}&{$\rm 2p_{1/2}$}&{203.4}&{\SI{6.25e+16}{}}&{\SI{7.38e3}{}}\\
{$^{93}$Mo}&{E2}&{1477.2}&{$\rm 2p_{3/2}$}&{228.53}&{\SI{1.21e+17}{}}&{\SI{1.27e4}{}}\\
{$^{138}$Ba}&{E1}&{6244.8}&{$\rm 1s_{1/2}$}&{44.19}&{\SI{2.69e+14}{}}&{\SI{146}{}}\\
{$^{202}$Hg}&{E1}&{4922}&{$\rm 2p_{1/2}$}&{329.41}&{\SI{6.12e+14}{}}&{\SI{44.66}{}}\\
{$^{207}$Pb}&{E1}&{4980.5}&{$\rm 2p_{1/2}$}&{165.51}&{\SI{3.09e+15}{}}&{\SI{447.50}{}}\\
{$^{207}$Pb}&{E1}&{4980.5}&{$\rm 2p_{3/2}$}&{350.64}&{\SI{5.67e+15}{}}&{\SI{388.75}{}}\\
\end{tabular}
\end{threeparttable}
\end{table*}
\noindent

In the table  \ref{tab:NEECvsNEUC}  we compare the NE\textmu C and NEEC resonance strengths of few of the strongest transitions. For all considered cases the NE\textmu C is substantially stronger than NEEC.  The enhancement found ranges between 5 to 10 order of magnitude. Table \ref{tab:NEECvsNEUC} offers also a comparison with the direct process of photoexcitation. Results show that in the case of an E1 transition, as for $^{138}$Ba and $^{207}$Pb, $S_\gamma$ and $S_{\rm NE\text{\textmu} C}$ are comparable, while for quadrupolar excitations $S_{\rm NE\text{\textmu} C}$ is substantially larger than $S_\gamma$.

\begin{table*}[!hbtp]
\centering
\caption{\label{tab:NEECvsNEUC} 
Comparison between NE\textmu C, NEEC and direct photo-excitation for the same nuclear transition for several isotopes. The apex \textit{i} indicates a bare nucleus configuration while \textit{n} the one for a neutral atom. Integrated cross sections are expressed in b$\cdot$eV, while $E_\mathrm{n}$ in \SI{}{\kilo\electronvolt}.
} 
\begin{tabular}{S|S|S|S|S|S|S} \toprule
{Isotope}&{$E_{\mathrm{n}}$}&{$nl_j$}&{${S^{\rm i}_{\mathrm{NE\text{\textmu} C}}}$}&{${S^{\rm i}_{\mathrm{NEEC}}}$}&{${S_{\mathrm{\gamma}}}$}&{${S^{\mathrm{n}}_{\mathrm{NE\text{\textmu} C}}}$}\\ \midrule 
{$^{52}$Mn}&{731.66}&{$\rm 2p_{1/2}$}&{\SI{771.14}{}}&{\SI{6.83e-7}{}}&{0.58}&{\SI{764.73}{}}\\
{$^{68}$Se}&{853.75}&{$\rm 2p_{3/2}$}&{\SI{1.82e5}{}}&{\SI{1.44e-5}{}}&{\SI{4.29}{}}&{\SI{1.62e5}{}}\\
{$^{73}$Ge}&{825.8}&{$\rm 2p_{1/2}$}&{\SI{9.85e3}{}}&{\SI{4.35e-6}{}}&{1.34}&{\SI{9.45e3}{}}\\
{$^{93}$Mo}&{1477.2}&{$\rm 2p_{3/2}$}&{\SI{1.27e4}{}}&{\SI{8.91e-6}{}}&{\SI{5.00}{}}&{\SI{1.24e4}{}}\\
{$^{138}$Ba}&{6244.8}&{$\rm 1s_{1/2}$}&{146}&{\SI{1.57e-2}{}}&{\SI{164.74}{}}&{\SI{120.66}{}}\\
{$^{207}$Pb}&{4980.5}&{$\rm 2p_{1/2}$}&{447.5}&{\SI{6.52e-2}{}}&{\SI{713.96}{}}&{\SI{432.6}{}}\\
\end{tabular}
\end{table*}
\noindent

\begin{figure}[!htbp]
\includegraphics[width=\linewidth]{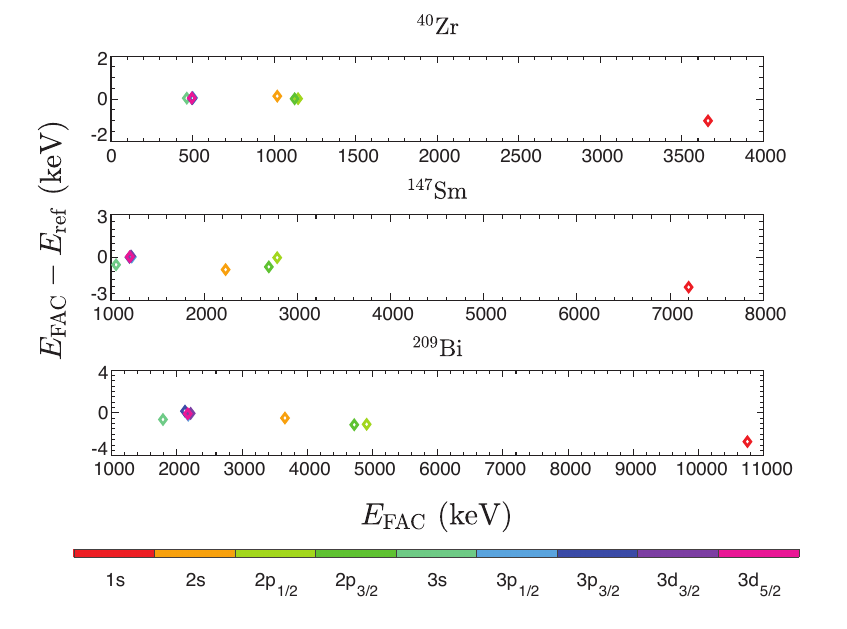}
\caption{Accuracy assessment of the Flexible Atomic Code. Muonic binding energies computed with FAC ($E_\mathrm{FAC}$) are compared with those of Ref. \cite{michel2017theoretical} ($E_\mathrm{ref}$). The color of the marker indicates the muonic state.}
\label{fig:Accuracy_M_FAC}
\end{figure}

Similarly to the NEEC case, the NE\textmu C cross section is greatly enhanced if the resonance is met at low kinetic energy, given the $\lambda^2$ prefactor in Eq. \ref{eq:NEEC_CS}. In this respect it is important to inspect the precision of atomic orbital calculations. Considering that FAC has never been used before to compute muonic binding energies, we compare in Fig. \ref{fig:Accuracy_M_FAC} the values obtained using FAC with the state-of-the-art theoretical calculations for muonic atoms presented in Ref. \cite{michel2017theoretical},  in the case of $^{40}$Zr, $^{147}$Sm and $^{209}$Bi. 
 
\noindent

The overall standard deviations between the differences in the binding energies range from \SI{0.36}{\kilo\electronvolt} for $^{40}$Zr to \SI{0.87}{\kilo\electronvolt} for $^{209}$Bi, assessing FAC as a valuable tool for predicting binding energies in muonic atoms (see Methods for further details). 

Another important difference with respect to NEEC is the absence of the high ionization state requirement. In case of muons, the muonic inner-shells are always available for capture and cannot be filled with electrons even for neutral atoms. The presence of electrons in the atomic environment will screen the muonic levels, making the muons less bound by  up to few tens of \SI{}{\kilo\electronvolt}, depending on the number of electrons in the shells \cite{vogel1973electron,fricke1969screening,michel2017theoretical}. This means that the resonance strengths  evaluated for bare nuclei in Table \ref{tab:Calculations} will be only slightly affected by the electronic charge state of the capturing ion. Thus, NE\textmu C allows for a capture in the $\rm 1s$ shell of an entirely filled atom. For this reason, we evaluated the NE\textmu C resonance strengths for the isotopes with the highest $S_\mathrm{NE\text{\textmu} C}$ of Table \ref{tab:Calculations} also in case of a neutral electronic configuration (see Methods for further details). Results are shown in Table \ref{tab:NEECvsNEUC}. Here we notice, as expected, that the ${S^{\mathrm{i}}_{\mathrm{NE\text{\textmu} C}}}$ and ${S^{\mathrm{n}}_{\mathrm{NE\text{\textmu} C}}}$ are very close to each other, with a slight difference due to the different resonance energy of the neutral case induced by the electron screening.

From the experimental point of view, the possibility of capture in neutral atoms can be tremendously useful and could offer an interesting perspective lifting the stringent experimental requirements for NEEC. Indeed, as NEEC simultaneously requires high ionization state and high density of resonant electrons, the simultaneous realization of both poses experimental challenges. Lifting the ionization requirement for NE\textmu C simplifies the experimental scenario.
For example, in a beam-based setup NE\textmu C can be observed by sending a muon beam into a solid target. 
Analogously to NEEC, The NE\textmu C probability can be written as \cite{aikawa2015thick,Wu2019}:

\begin{equation}
P=\sum_{\alpha_{r}} n_i\ S_\mathrm{NE \text{\textmu} C}^{\alpha_\mathrm{r}}\ \dfrac{1}{-(dE_\text{\textmu} /dx) \vert_{E_\mathrm{r}}}
\end{equation}
\noindent
where $\alpha_\mathrm{r}$ represent the available capture channels, $n_\mathrm{i}$ is the density of atoms and $-(dE_\text{\textmu} /dx) \vert_{E_\mathrm{r}}$ is the muon stopping power at the resonance energy .
The number of excited nuclei per second, assuming a continuous muon beam with a flux $\phi_\text{\textmu}$ (1/s) is given by:
\begin{equation}
N_\mathrm{NE\text{\textmu} C}^{\mathrm{exc}} = P \phi_\text{\textmu} .
\end{equation}
\noindent
If we limit ourselves to solid targets with stable or long-lived ground states, essential for practical experiments, $^{73}$Ge and $^{93}$Mo are the most promising isotopes. The stopping power calculated with GEANT4  \cite{agostinelli2003geant4} is of $dE_\text{\textmu} /dx \simeq \SI{-501}{\mega\electronvolt\per\centi\meter}$ and  \SI{-607}{\mega\electronvolt\per\centi\meter} at the resonant energies of \SI{101.84}{\kilo\electronvolt} and \SI{228.53}{\kilo\electronvolt} respectively. Considering the capture only in the $\rm 2p_{3/2}$ channel the resulting probabilities are $P=\SI{1.54e-6}{}$ and $P=\SI{1.39e-6}{}$ respectively. Remarkably, these theoretical probabilities are  five orders of magnitude larger than those theoretically estimated for the $^{93m}$Mo isomer depletion through NEEC \cite{Wu2019,rzadkiewicz2021novel}. If we expand the calculations to short lived isotopes, e.g. $^{68}$Se , the single channel excitation probability reaches $P=\SI{1.02e-5}{}$. 

Currently, the brightest $\text{\textmu}^{-}$ beam facilities at PSI (Villigen, Switzerland) and MuSIC (Osaka, Japan) are able to deliver a continuous flux of $10^{7}$ muons per second \cite{cook2017delivering}. Planned upgrades would make it feasible in the next years to have fluxes up to $10^{8}$ and $10^{9}$ muons per second, resulting approximately in ten to one thousand nuclear excitation per second.
Futhermore, an increase in the excitation cross section is expected if the wavefunction of the muon is engineered \cite{vanacore2020spatio}, i.e. considering muon vortex beams \cite{zhao2021decay}, as recently suggested for NEEC \cite{Madan2020,gargiulo2021nuclear,wu2021dynamical}. This modification of the wavefunction could make unfavourable transitions with higher multipolarity more likely to happen.

Given the high energy of nuclear transitions involved in NE\textmu C and its increased efficiency compared to direct photoexcitation at higher multipolarities, NE\textmu C can be the most suitable process for isomer feeding. In this case the feeding, as shown in Fig. \ref{fig:Perspective} will not happen directly to the isomer state, but arriving to it through subsequent decays upon the initial excitation from the ground state. This is, for example, the case of the energy level schemes of $^{113}$In and $^{87}$Sr. 

Typically, at energies tens of \SI{}{\mega\electronvolt} above the ground state, the density of the excitation states is so high that they overlap in a broad energy range, giving rise to the so called giant resonances. Excitation of these resonances, independently from the particular excitation mechanism, can lead to fission if the resonance is above the fission barrier. 
Prompt fission of the nucleus has been achieved under muon excitation and attributed to the muon cascade in $^{238}$U \cite{oberacker1993muon}. Yet, the possibility of the NE\textmu C has not been considered despite it could provide substantially better energy overlap given that one has an additional degree of freedom, that is the energy of the free lepton.
Indeed, fission induced by muonic transitions is governed by the energy difference between two muonic bound states, while in the case of NE\textmu C the resonance condition is satisfied throughout the whole width of the giant resonance, that can be several \SI{}{\mega\electronvolt} wide. 
To estimate the contribution of the NE\textmu C process into the muon induced fission 
we calculate the fission cross section induced by NE\textmu C for $^{238}$U presented in the Extended Data Figure \ref{fig:NEuC_fission}  (see Methods for further details). Integrating the cross section with the energy dependent stopping power (further details are available in the Methods), provides us with final fission probability of $\sim \SI{4.30e-5}{}$ per incident muon. This probability is still small if compared with prompt fission induced by muon cascade ($\sim \SI{e-3}{}$) and delayed fission induced by muon capture ($\sim \SI{e-2}{}-\SI{e-1}{}$) \cite{johansson1980muon}.
Nevertheless, for lighter isotopes the muon cascade eventually becomes non-resonant with the giant resonances, while NE\textmu C is theoretically always possible. 

Most remarkably, the  NE\textmu C has the highest chance to be observed than the NEEC process in which disagreement between experiment and theory is of nine orders of magnitude. Measuring the NE\textmu C rates and comparing to the estimates provided in presented paper will hopefully help to resolve the contradiction and establish the origins of the extremely high experimentally measured NEEC cross sections.  

\bmhead{Acknowledgments} S.G., F.C. and I.M. acknowledge support from Google Inc.
The authors would like to thank Giovanni dal Maso for the useful insights about muonic beam facilities.

\section*{Methods}\label{sec11}
\subsection*{Binding energy calculations in muonic atoms using the Flexible Atomic Code}
Binding energies for muonic atoms are obtained by numerically solving the Dirac equation including the effect of the finite size of the nucleus using the Fermi distribution function with parameters adjusted to reproduce the rms charge radii of Ref. \cite{angeli2013table}. Vacuum polarization is taken into account using the standard Uehling potential, while self-energy correction is included using the method of Ref. \cite{barrett1968}. The nucleus recoil effect is approximated with an effective Hamiltonian term proposed in Ref. \cite{cheng1991}. As we can see from Fig. \ref{fig:Accuracy_M_FAC}, maximum discrepancy is obtained for the $\rm 1s$ shell with differences ranging between \SI{1.08}{\kilo\electronvolt} for $^{40}$Zr and \SI{2.81}{\kilo\electronvolt} for $^{209}$Bi. In the case of the $\rm 2s$ subshell, discrepancies drops down to a minimum of \SI{0.07}{\kilo\electronvolt} for $^{40}$Zr to a maximum of \SI{0.94}{\kilo\electronvolt} for $^{147}$Sm. For the $\rm 2p_{1/2}$ the maximum disagreement, of about  \SI{1.19}{\kilo\electronvolt}, results from $^{209}$Bi. Binding energies for the $\rm 2p_{3/2}$ subshell similarly differ at most by \SI{1.24}{\kilo\electronvolt} in the case of $^{209}$Bi, while the discrepancies are smaller for $^{40}$Zr and $^{147}$Sm. Much of these discrepacies can be attributed to the self-energy correction included in the present work, and omitted in Ref. \cite{michel2017theoretical}. The agreement between FAC and Ref. \cite{michel2017theoretical} improves significantly for the M-shell, as the self-energy term becomes negligible. In this context we used the rms nuclear charge radii reported in Ref. \cite{angeli2013table} and $t=\SI{2.30}{\femto\meter}$ as the parameter of the Fermi-type charge distribution. Screening by an arbitrary electronic configuration has been included in FAC by solving the Dirac equations of both the muon and electrons self consistently via iteration.

\subsection*{Radial integral calculation}
The radial integral $\widetilde{R}_{L,k_b,k}$ has been calculated using its full expression, reported in Ref. \cite{Gagyi-Palffy06}, that is:

\begin{eqnarray}
    \widetilde{R}_{L,k_b,k} &=& \dfrac{1}{R_0^{L-1}}\int_0^{R_0} dr\ r^{L+2} [f_{k_b}(r)f_{E k}(r) + g_{k_b}(r)g_{E k}(r) ]+ \nonumber \\ 
 & & R_0^{L+2} \int_{R_0}^{+\infty} dr\ r^{-L+1} [f_{k_b}(r)f_{E k}(r) + g_{k_b}(r)g_{E k}(r) ]\ ,
\end{eqnarray}
\noindent
where $f(r)$ and g(r) are the large and small radial components of the bound and free muon wavefunctions. $R_0$ is the nuclear radius for which we used the values reported in Ref. \cite{angeli2013table} for the available isotopes. For those not available we used the value reported by FAC, that is \SI{3.56208}{\femto\meter} for $^{48}$Cr, \SI{3.96475}{\femto\meter} for $^{68}$Se and \SI{4.36807}{\femto\meter} for $^{93}$Mo.
\subsection*{NE\textmu C-induced fission cross section}
$^{238}$U presents the giant dipole resonance peak at $E_{\rm GDR}=\SI{12.8}{\mega\electronvolt}$, that is slightly above the muonic binding energy of the K-shell, $E_{\rm b}^\mathrm{K}=\SI{12.12}{\mega\electronvolt}$. As similarly shown in Ref. \cite{zaretski1961theory}, the NE\textmu C-induced fission has been related to the value of the photofission cross section through the matrix element of the dipole transition, thus to the photoexcitation cross section. 
Here, we use Eq. (XII, 7.27) in Ref. \cite{blatt1991theoretical} as an expression for the direct photo-excitation cross section in case of high-energy nuclear transitions. Since, under Bohr assumption, the decay mode of the excited nucleus is independent from its formation, we express the relation between between NE\textmu C-induced fission cross section and the photo-fission cross section as:

\begin{eqnarray}
 \sigma_{\rm NE\text{\textmu} C}(\text{\textmu}, F) &=& \dfrac{1}{4\pi\alpha} \dfrac{4\pi^2 \rho_i}{(2L+1)^2 } (2 j_b +1)\nonumber \times \sum_k \lvert\widetilde{R}_{L,k_b,k}\rvert^2 C(j_b\ L\ j; 1/2\ 0\ 1/2)^2\ \\
 & &\times  2\pi^2 \lambda_\mathrm{\text{\textmu}}^2 \dfrac{3}{8\pi^3} \dfrac{1}{E} \sigma_{\gamma} (\gamma, F).
\end{eqnarray}
where $E$ is the energy of the transition. Photo-fission cross section has been retrieved from Fig. 4 of Ref. \cite{gindler1956photofission} and is also shown for comparison in the Extended Data Figure \ref{fig:NEuC_fission}. 
The radial integral $\widetilde{R}_{L,k_b,k}$ has been calculated over the range \SI{100}{eV} to \SI{8}{\mega\electronvolt} for the free muon energy, considering a neutral electronic configuration for $^{238}$U. 
The probability of fission induced by the nuclear excitation by muon capture is then evaluated as:

\begin{equation}
P= n_i\ \int dE \dfrac{\sigma_\mathrm{NE \text{\textmu} C}(\text{\textmu},F)
}{-(dE_\text{\textmu} /dx)}
\label{eq:integrated_prob}
\end{equation}
\noindent
considering the only capture in the 1s shell. The stopping power is taken by Ref. \cite{groom2001muon}, neglecting the small difference between positive and negative muons.
The integral in Eq. \ref{eq:integrated_prob} has been performed choosing $E=\SI{100}{\electronvolt}$ as lower limit for the integration, that is the minimum energy for which the stopping power is available.

\clearpage
\section*{Extended Data Figures}
\renewcommand{\figurename}{Extended Data Figure}
\setcounter{figure}{0}
\begin{figure}[!htbp]
\includegraphics[width=\linewidth]{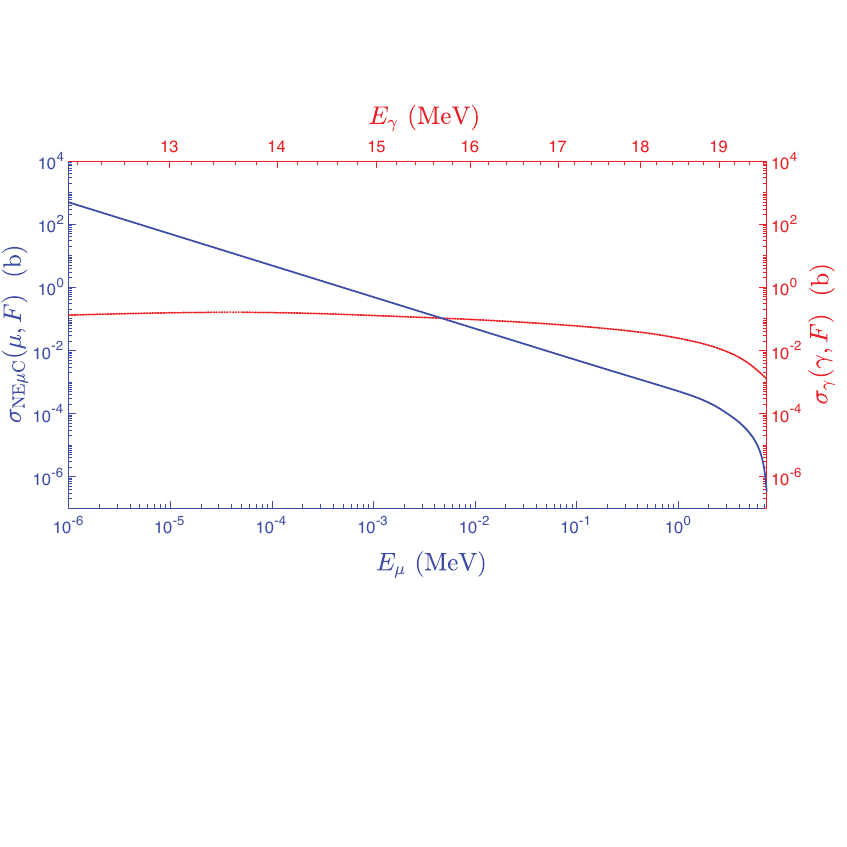}
\caption{NE\textmu C-induced fission cross section as function of the free muon kinetic energy $E_{\mu}$ (blue axes and curve) and photo-fission cross section (red axes and curve) retrieved from Ref. \cite{gindler1956photofission} for $^{238}$U. X-axes are aligned considering the relation between the energy of the free muon and the nuclear transtion, i.e. $E_\mu=E_\mathrm{n}-E_\mathrm{b}^\mathrm{K}$. In this case, $E_\gamma=E_\mathrm{n}$ and $E_\mathrm{b}^\mathrm{K}=\SI{12.12}{\mega\electronvolt}$.}
\label{fig:NEuC_fission}
\end{figure}

\noindent

\end{document}